\documentstyle[epsbox,12pt]{article}

\begin{document}

\newcommand{\dfrac}[2]{\displaystyle{\frac{#1}{#2}}}

{\it University of Shizuoka}

\hspace*{10cm}{\bf US-99-08}\\[-.3in]

%\hspace*{10cm}{\bf hep-ph/9809548}\\[-.3in]

\hspace*{10cm}{\bf September 1999}\\[.4in]

\begin{center}

{\Large\bf SO(10)$\times$SO(10) Universal Seesaw Model\\[.2in]
and its Intermediate Mass Scales}

{\bf Yoshio Koide}\footnote{E-mail address: koide@u-shizuoka-ken.ac.jp}
\\[.1in]

{ Department of Physics, University of Shizuoka} \\

{ 52-1 Yada, Shizuoka 422-8526, Japan}

%\date{\today}
\vspace{2cm}

{\large\bf Abstract}

\end{center}

\begin{quotation}
On the basis of the universal seesaw mass matrix model, 
which is a promising model of the unified description of the
quark and lepton mass matrices, the behaviors of
the gauge coupling constants and intermediate 
energy scales in the ${\rm SO(10)}_L\times{\rm SO(10)}_R$ 
model are investigated related to the neutrino mass
generation scenarios.
The non-SUSY model cannot give favorable values of the
intermediate energy scales to explain the smallness 
of the neutrino masses, while the SUSY model can give the 
plausible values if the number $n_\phi$ of the weak doublet 
Higgs scalars is $n_\phi \geq 3$.
\end{quotation}

\vfill
{Key words: universal seesaw, evolution, SO(10), 
neutrino mass matrix, 

intermediate energy scale}


To be published in Physical Review D.
%%%%%%%%%%%%%%%%%%%%%%%%%%%%%%%%%%%%%%%%%%%
\newpage

\section{Introduction}
\label{sec:1}

%\vglue.1in

Recently, a universal seesaw mass matrix model has 
considerably attracted us as a unified mass matrix 
model of the quarks and leptons. 
The model \cite{USMM} was proposed 
in order to understand the question why the masses of 
quarks (except for top quark) and charged leptons are
so small compared with the electroweak scale
$\Lambda_L$ ($\sim$ 10$^2$ GeV).
The model has hypothetical fermions
$F_i$ in addition to the conventional quarks
and leptons $f_i$ (flavors $f=u, d, \nu, e$; family
indices $i=1,2,3$), and those are assigned to 
$f_L$ = (2,1), $f_R$ = (1,2), $F_L$ = (1,1) and
$F_R$ = (1,1) of SU(2)$_L \times$ SU(2)$_R$.
The 6 $\times$ 6 mass matrix which is sandwiched
between the fields ($\overline{f}_L, \overline{F}_L$)
and ($f_R, F_R$) is given by
$$
M^{6 \times 6} =
\left( \begin{array}{cc}
0 & m_L\\
m_R & M_F
\end{array} \right) \ ,
\eqno(1.1)
$$
where $m_L$ and $m_R$ are universal for all fermion
sectors ($f=u, d, \nu, e$) and only $M_F$ have
structures dependent on the flavors $f$.
For $\Lambda_L < \Lambda_R \ll \Lambda_F$, where 
$\Lambda_L = O(m_L), \Lambda_R = O(m_R)$ and
$\Lambda_F = O(M_F)$, the $3\times 3$ mass matrix
$M_f$ for the fermions $f$ is given by the
well-known seesaw expression $M_f \simeq - m_L M^{-1}_F m_R$,
so that the quarks and lepton masses $m_{q,l}$ are given with
a suppression factor $\Lambda_R/\Lambda_F$.  
In order to understand the observed  heavy top quark mass value 
$m_t \sim \Lambda_L$, we put an additional condition 
{\rm det}$M_F = 0$ on the up-quark sector ($F=U$) 
\cite{KFzp,Morozumi}.
Then, since one of the fermion masses $m(U_i)$ ($i=1,2,3$) is zero 
[say, $m(U_3)=0$], so that the seesaw mechanism does not work for
the third family, and the fermions ($u_{3L}, U_{3R}$)
and ($U_{3L}, u_{3R}$) acquire masses of $O(m_L)$ and
$O(m_R)$, respectively, without the suppression factor 
$\Lambda_R/\Lambda_F$. 
We identify $(u_{3L},U_{3_R})$ as the top quark 
$(t_L, t_R)$.
An explicit model for the matrix forms $m_L$, $m_R$ and 
$M_F$ has been proposed by Fusaoka and the author \cite{KFzp},
and they have successfully obtained the numerical results on
the quark masses and mixings, where those quantities are 
described in terms of the charged lepton masses by assuming 
simple structures of $m_L$, $m_R$ and $M_F$.

For the neutrino mass matrix, we start the following $12\times 12$
mass matrix:
$$
\left( \overline{\nu}_L\ \overline{\nu}_R^c\ \overline{N}_L\ 
\overline{N}_R^c \right)
{ 
\left( \begin{array}{cccc}
0 & 0 & 0 & m_L \\
0 & 0 & m_R^T & 0 \\
0 & m_R & M_R & M_D \\
m_L^T & 0 & M_D^T & M_L \\
\end{array} \right)
}
\left( \begin{array}{c}
\nu_L^c \\
\nu_R \\
N_L^c \\
N_R
\end{array} \right) \ .
\eqno(1.2)
$$
The mass matrix (1.2) leads to different scenarios of the 
neutrino phenomenology correspondingly to the different 
structure of the intermediate mass scales 
$\Lambda_{NL}=O(M_L)$, $\Lambda_{NR}=O(M_R)$, 
and $\Lambda_{D}=O(M_D)$ together with to $\Lambda_L$, $\Lambda_R$, 
and $\Lambda_F$ ($F\neq N$).
For example, for the case 
$\Lambda_{NL}, \Lambda_{NL}\geq \Lambda_{D}$, 
the neutrino mass matrix is approximately given by 
$M_\nu \simeq - m_L M_L^{-1} m_L^T$, so that the neutrino 
masses are suppressed by the factors $\Lambda_L/\Lambda_R$ and 
$\Lambda_F/\Lambda_{NL}$ compared with the quark and charged 
lepton masses $m_{q,l}$.

In spite of such phenomenological successes, there is a
reluctance to recognize the model, because the model 
needs extra fermions $F$.
In most unification models, there are no rooms for the 
fermions $F$.
Whether we can built a unification model in which
the fermions $F$ are reasonably embedded will be 
a touchstone for the great future of the universal 
seesaw mass matrix model. 

For this problem, there is an attractive idea \cite{so10}.
We can consider that the fermions $F_R^c$ 
($\equiv C\overline{F}_R^T$)
together with the fermions $f_L$ belong to {\bf 16} of SO(10), and 
also $F_L^c$ together with $f_R$ belong to {\bf 16} of another
SO(10), i.e.,
$$
(f_L + F_R^c) \sim (16,1)\ , \ \ \ (f_R + F_L^c) \sim (1,16) \ , 
\eqno(1.3)
$$
of SO(10)$_L\times$SO(10)$_R$.
In order to examine the idea (1.3), in the present paper,
we investigate the evolution of the gauge coupling constants 
on the basis of the SO(10)$_L\times$SO(10)$_R$ model and 
estimate the intermediate energy scales $\Lambda_R$, $\Lambda_F$ 
and $\Lambda_N$ together with the unification energy scale $\Lambda_X$. 
The evolutions of the gauge coupling constants under
SO(10)$_L\times$SO(10)$_R$ symmetries have already been done 
by Davidson, Wali and Cho \cite{so10}.
The case of the symmetry breaking ${\rm SO(10)}_L\times 
{\rm SO(10)}_R \rightarrow [{\rm SU(5)}\times {\rm U(1)}]_L 
\times [{\rm SU(5)}\times {\rm U(1)}]_R$ is easily ruled 
out phenomenologically.
On the contrary, it is not clear that the case 
${\rm SO(10)}_L\times {\rm SO(10)}_R \rightarrow 
[{\rm SU(2)}\times {\rm SU(2)}' \times {\rm SU(4)}]_L \times 
 [{\rm SU(2)}\times {\rm SU(2)}' \times {\rm SU(4)}]_R$ is 
ruled out or not,
because there are many symmetry breaking patterns which were not 
discussed in the Ref.~\cite{so10}. 
In the present paper, we will systematically investigate 
the intermediate mass scales for all possible cases (including non-SUSY
and SUSY cases), but under a numerical constraint
$\Lambda_R / \Lambda_F \simeq 0.02$ \cite{KFzp} which were derived 
from the observed ratio $m_c/m_t$ under the new scenario 
of the universal seesaw model.

In Sec.~\ref{sec:2}, we introduce the Higgs scalars in the present model
and classify the possible cases of the symmetry breaking patterns.
In Sec.~\ref{sec:3}, we shortly review possible forms of the active 
neutrinos $\nu_L$ under the various cases of the intermediate 
energy scales.
In Sec.~\ref{sec:4}, we discuss the evolution of the gauge coupling 
constants for the case ${\rm SO(10)}_L \times {\rm SO(10)}_R \rightarrow 
[{\rm SU(2)}\times {\rm SU(2)}'\times {\rm SU(4)}]_L \times 
[{\rm SU(2)}\times {\rm SU(2)}'\times {\rm SU(4)}]_R$.
The numerical results are presented in Sec.~\ref{sec:5}.
Although non-SUSY cases give the value 
$\Lambda_L/\Lambda_R \sim 10^{-5}$, the value is not sufficient
to explain the smallness of the neutrino masses.
On the other hand, the SUSY case can give a reasonable order of
$\Lambda_L/\Lambda_R \sim 10^{-10}$ for $n_\phi=6$, where $n_\phi$
is the number of the SU(2) doublet Higgs scalars.  However, the 
model will encounter a new problem, i.e., the flavor-changing
neutral currents (FCNC) problem.
Sec.~\ref{sec:6} will be devoted to the conclusions and 
remarks.

%%%%%%%%%%%%%%%%%%%%%%%%%%%%%%%%%%%%%%%%%%%%%%%%%%%%%%%%%%%%%%%%%
\vspace{.2in}

\section{Higgs bosons and possible symmetry breaking patterns}
\label{sec:2}

%\vglue.1in

In the present model, we consider the following Higgs scalars:

\noindent (i) $\Phi_{XL}=(54,1)$ and $\Phi_{XR}=(1,54)$, whose 
vacuum expectation values (VEV) break  the symmetries 
${\rm SO(10)}_L$ and ${\rm SO(10)}_R$ into  
$ [{\rm SU(2)} \times {\rm SU(2)}' \times {\rm SU(4)}]_L$ and 
$[{\rm SU(2)} \times {\rm SU(2)}' \times {\rm SU(4)}]_R$, 
respectively;

\noindent (ii) $\Phi_{NL}=(126,1)$ and $\Phi_{NR}=(1,126)$,
whose VEV break the symmetries $[{\rm SU(2)}' \times {\rm SU(4)}]_L$
and  $[{\rm SU(2)}' \times {\rm SU(4)}]_R$ into 
$[{\rm U(1)} \times {\rm SU(3)}]_L$ and 
$[{\rm U(1)} \times {\rm SU(3)}]_R$, respectively, and 
generate the Majorana mass terms $\overline{N}_R^c M_L N_R$
and $\overline{N}_L M_R N_L^c$, respectively;

\noindent (iii)  $\Phi_{F}=(\overline{16},16)$, whose VEV
break ${\rm SU(3)}_L \times {\rm SU(3)}_R$ into  ${\rm SU(3)}_{LR}$
and ${\rm U(1)}_L \times {\rm U(1)}_R$ into  ${\rm U(1)}_{LR}$,
and generate the Dirac mass terms  $\overline{F}_L M_F F_R$;

\noindent (iv) $\phi_{L}=(10,1)$ and $\phi_{R}=(1,10)$, 
whose VEV break  SU(2)$_L$ and SU(2)$_R$ and generate 
the mass terms $\overline{f}_L m_L F_R$ and 
$\overline{F}_L m_R f_R$, respectively.

For example, we consider the following case of the 
symmetry breaking pattern:

\noindent
Case RLRL:  $\Lambda_{XR} > \Lambda_{XL} >
\Lambda_{NR} > \Lambda_{NL} = \Lambda_D > \Lambda_{F}$ 
$$
{\rm SO(10)}_L \times {\rm SO(10)}_R
$$
$$
\downarrow \ \ \  {\rm at}\ { \mu=\Lambda_{XR}}
$$
$$
\rm SO(10)_L \times 
[{\rm SU(2)}\times{\rm SU(2)'}\times{\rm SU(4)}]_R
$$
$$
\downarrow \ \ \  {\rm at}\ { \mu=\Lambda_{XL}}
$$
$$
[{\rm SU(2)}\times{\rm SU(2)'}\times{\rm SU(4)}]_L \times 
[{\rm SU(2)}\times{\rm SU(2)'}\times{\rm SU(4)}]_R
$$
$$
\downarrow \ \ \  {\rm at}\ { \mu=\Lambda_{NR}}
$$
$$
[{\rm SU(2)}\times{\rm SU(2)'}\times{\rm SU(4)}]_L \times 
[{\rm SU(2)}\times{\rm U(1)}\times{\rm SU(3)}]_R
$$
$$
\downarrow \ \ \  {\rm at}\  \mu=\Lambda_{NL}=\Lambda_D
$$
$$
[{\rm SU(2)}\times{\rm U(1)}\times{\rm SU(3)}]_L \times 
[{\rm SU(2)}\times{\rm U(1)}\times{\rm SU(3)}]_R
$$
$$
\downarrow \ \ \  {\rm at}\  \mu=\Lambda_{F}
$$
$$
{\rm SU(2)}_L\times {\rm SU(2)}_R\times {\rm U(1)}_{LR} 
\times {\rm SU(3)}_{LR}  
$$
$$
\downarrow \ \ \  {\rm at}\  \mu=\Lambda_{R}
$$
$$
{\rm SU(2)}_L\times {\rm U(1)}_{Y} \times {\rm SU(3)}_{LR} 
$$
$$
\downarrow \ \ \  {\rm at}\  \mu=\Lambda_{L}
$$
$$
{\rm U(1)}_{em} \times  {\rm SU(3)}_{LR} 
 \ ,
\eqno(2.1)
$$
where SU(3)$_{LR}$ means the color gauge symmetry SU(3)$_c$.

For convenience, we define the following ranges of 
the energy scale $\mu$:
Range 1 ($\Lambda_L < \mu \leq \Lambda_R$);
Range 2 ($\Lambda_R < \mu \leq \Lambda_F$);
Range 3 ($\Lambda_F < \mu \leq \Lambda_{NL}$);
Range 4 ($\Lambda_{NL} < \mu \leq \Lambda_{NR}$);
Range 5 ($\Lambda_{NR} < \mu \leq \Lambda_{XL}$);
Range 6 ($\Lambda_{XL} < \mu \leq \Lambda_{XR}$).
Here, we regard $\Lambda_D=\Lambda_{NL}$. 
Also, for convenience of the next section, let us define 
the following parameters
$$
x_1=\log(\Lambda_R/\Lambda_L) \ , \ \ 
x_2=\log(\Lambda_{F}/\Lambda_R) \ , \ \ 
x_3=\log(\Lambda_{NL}/\Lambda_F) \ , 
$$
$$  
x_4=\log(\Lambda_{NR}/\Lambda_{NL})\ , \ \ 
x_5=\log(\Lambda_{XL}/\Lambda_{NR})\ , \ \ 
x_6=\log(\Lambda_{XR}/\Lambda_{XL})\ . 
\eqno(2.2)
$$
The values of $x_i$ must be positive or zero.
Especially, the value of $x_1$ must roughly be $x_1 \geq 1$
from the experimental lower limit \cite{pdg98} of the 
right-handed weak boson mass.

We also investigate the following cases:

\noindent
Case RLLR:  $\Lambda_{XR} > \Lambda_{XL} > \Lambda_{NL} > 
\Lambda_{NR} = \Lambda_D > \Lambda_{F}$; 

\noindent
Case LRLR:  $\Lambda_{XL} > \Lambda_{XR} >
\Lambda_{NL} > \Lambda_{NR} = \Lambda_D > \Lambda_{F}$; 

\noindent
Case LRRL:  $\Lambda_{XL} > \Lambda_{XR} >
\Lambda_{NR} > \Lambda_{NL} = \Lambda_D > \Lambda_{F}$; 

\noindent
Case RRLL:  $\Lambda_{XR} > \Lambda_{NR} >
\Lambda_{XL} > \Lambda_{NL} = \Lambda_D > \Lambda_{F}$; 

\noindent Case
LLRR:  $\Lambda_{XL} > \Lambda_{NL} >
\Lambda_{XR} > \Lambda_{NR} = \Lambda_D > \Lambda_{F}$. 

We consider a model without $\Phi_{NL}$:

\noindent
Case RLLD:  $\Lambda_{XR} > \Lambda_{XL} >
\Lambda_{NL} > \Lambda_{D} > \Lambda_{F}$; 

\noindent
Case  LRLD:  $\Lambda_{XL} > \Lambda_{XR} >
\Lambda_{NL} > \Lambda_{D} > \Lambda_{F}$;

\noindent a model without $\Phi_{NL}$:

\noindent
Case  RLRD:  $\Lambda_{XR} > \Lambda_{XL} >
\Lambda_{NR} > \Lambda_{D} > \Lambda_{F}$;

\noindent
Case  LRRD:  $\Lambda_{XL} > \Lambda_{XR} >
\Lambda_{NR} > \Lambda_{D} > \Lambda_{F}$;

\noindent
and a model without $\Phi_{NL}$ and $\Phi_{NR}$

\noindent
Case  RLD:  $\Lambda_{XR} > \Lambda_{XL} >
 \Lambda_{D} > \Lambda_{F}$;

\noindent
Case  LRD:  $\Lambda_{XL} > \Lambda_{XR} >
 \Lambda_{D} > \Lambda_{F}$.

In addition to these cases, we investigation the following
cases, which lead to a model with pseudo-Dirac neutrino states
which was pointed out by Bowes and Volkas \cite {Bowes}:

\noindent
Case RLDN:  $\Lambda_{XR} > \Lambda_{XL} >
\Lambda_D > \Lambda_F > \Lambda_{NR/NL}$;

\noindent
Case  LRDN:  $\Lambda_{XL} > \Lambda_{XR} >
\Lambda_D > \Lambda_F > \Lambda_{NR/NL}$.

For each case, in a similar way to the case RLRL, we define 
the energy scale regions and parameters $x_i$.
The definitions for some typical cases are listed in Table
\ref{T-sym}.
For the model without the scalars $\Phi_{NL}$ and 
$\Phi_{NR}$ (the cases RLD and LED) and the Bowes-Volkas model 
(the cases RLDN and LRDN),
we define the ranges without the range 4 (the parameter $x_4$).
The definitions for the other cases which are not given in
Table \ref{T-sym} can readily be read by the 
exchange $L\leftrightarrow R$.

Each case is investigated for the cases of non-SUSY and SUSY.
Here, the ``SUSY" case means a minimal SUSY model, and for simplicity,
we take the SUSY breaking energy scale $\Lambda_{SUSY}$
as $\Lambda_{SUSY}=\Lambda_L$ in the numerical estimates.

%%%%%%%%%%%%%%%%%%%%%%%%%%%%%%%%%%%%%%%%%%%%%%%%%%%%%%%%%%%%%%%%%
\vspace{.2in}

\section{Mass matrix for the active neutrinos}
\label{sec:3}

%\vglue.1in

Our interest is in the effective mass matrix $M_\nu$ for
the active neutrinos $\nu_L$.
For $\Lambda_D\gg \Lambda_R \gg \Lambda_L$, the mass matrix 
(1.2) approximately 
leads to the mass matrix for the neutrinos $(\nu_L^c, \nu_R)$:
$$
M^{6\times 6}\simeq - \left(
\begin{array}{cc}
m_L M_{22}^{-1}m_L^T & m_L M_{21}^{-1}m_R \\
m_L^T M_{12}^{-1}m_R^T & m_R^T M_{11}^{-1}m_R
\end{array} \right)  \ ,
\eqno(3.1)
$$
where
$$
\left(
\begin{array}{cc}
M_{11}^{-1} &  M_{12}^{-1} \\
M_{21}^{-1} &  M_{22}^{-1}
\end{array} \right) \ 
=\left(
\begin{array}{cc}
M_R &  M_D \\
M_D^T &  M_L
\end{array} \right)^{-1} \ , 
\eqno(3.2)
$$
$$
\begin{array}{l}
M_{11}=M_R -M_D M_L^{-1} M_D^T  \ ,\\
M_{22}=M_L -M_D^T M_R^{-1} M_D \ , \\
M_{12}=M_{21}^T=M_D^T -M_L M_D^{-1} M_R \ .\\
\end{array}
\eqno(3.3)
$$
The scenarios for neutrino masses and mixings are highly dependent on
the structure of the intermediate mass scales $\Lambda_{NL}$, 
$\Lambda_{NR}$, and $\Lambda_{D}$ relative to $\Lambda_{F}$ 
($F\neq N$), $\Lambda_{R}$, and $\Lambda_{L}$.
In this section, let us review  possible matrix forms of 
the effective mass matrix of the active neutrinos $\nu_L$.

\noindent
{\bf  Case A} : \ $M_L, M_R \geq M_D$ 

For the case $M_L, M_R \geq M_D$, the $6\times 6$ matrix (3.1) 
becomes
$$
M^{6\times 6}\simeq  \left(
\begin{array}{cc}
-m_L M_{L}^{-1}m_L^T & m_L M_L^{-1} M_D^T M_R^{-1} m_R \\
m_R^T M_R^{-1} M_D M_L^{-1}m_L^T & -m_R^T M_{R}^{-1} m_R
\end{array} \right) \ ,
\eqno(3.4)
$$
so that the approximate mass matrix $M(\nu_L)$ for the active 
neutrinos $\nu_L$ 
$$
M(\nu_L) \simeq -m_L M_{L}^{-1}m_L^T \ , 
\eqno(3.5)
$$
together with the mass matrix for the neutrinos $\nu_R$
$$
M(\nu_R) \simeq -m_R^T M_{R}^{-1}m_R \ . \\
\eqno(3.6)
$$
The smallness of the neutrino masses is given by
$$
m_\nu \sim \frac{\Lambda_L}{\Lambda_R} \cdot 
\frac{\Lambda_F}{\Lambda_{NL}} \cdot m_{e,q} \ .
\eqno(3.7)
$$

For the case $M_R \gg  M_L \sim M_D$, especially for the case 
$\Lambda_L^2/\Lambda_{NL}\sim \Lambda_R^2/\Lambda_{NR}$,
we can build an interesting scenario, where the solar
neutrino data \cite{solar-nu}, atmospheric neutrino 
data \cite{atm-nu} and LSND data \cite{LSND} are explained 
from a small mixing $\nu_{eL}\leftrightarrow \nu_{eR}^c$, 
a large mixing $\nu_{\mu L}\leftrightarrow \nu_{\tau L}$, and
a small mixing $\nu_{\mu L}\leftrightarrow \nu_{e L}$, respectively.
In order to realize the smallness of the neutrino masses,
$m_\nu/m_{e,q} \sim 10^{-9}$, 
the constraint 
$$
\Lambda_{NL} \Lambda_{NR}/\Lambda_F^2 \sim 10^{18} \ ,
\eqno(3.8)
$$
is required.

\noindent
{\bf  Case B} : \ $M_L, M_R \ll M_D$ 

For the case $M_R, M_L \ll M_D$, the matrix (3.1) leads to 
$$
M^{6\times 6}\simeq 
 \left(
\begin{array}{cc}
-m_L M_D^{-1} M_R M_D^{T-1} m_L^T & - m_L M_D^{-1} m_R \\
m_R^T M_D^T m_L^T & -m_R^T M_D^{T-1} M_L M_D^{-1} m_R
\end{array} \right) \ .
\eqno(3.9)
$$
Since $(M^{6\times 6})_{12}\gg (M^{6\times 6})_{11(22)}$
because of $m_R\gg m_L$, the case leads to a model with
the pseudo-Dirac neutrino states
$\nu_{i\pm} \simeq (\nu_{Li}\pm \nu_{Ri}^c)/\sqrt{2}$,
whose masses are given by the order
$$
m(\nu_\pm) \sim \Lambda_L\Lambda_R/\Lambda_D \sim 
(\lambda_F/\Lambda_D) m_{e,q} \ .
\eqno(3.10)
$$
This case has been discussed by Bowes and Volkas \cite{Bowes}.

\noindent
{\bf  Case C}: \ $M_R=0$

For a model without $\Phi_{NR}$, the matrix (3.1) leads to
$$
M^{6\times 6}\simeq  \left(
\begin{array}{cc}
0 & -m_L M_D^{-1} m_R \\
-m_R^T M_D^{T-1} m_L^T & m_R^T M_D^{T-1} M_L M_{D}^{-1} m_R
\end{array} \right) \ .
\eqno(3.11)
$$
When $(\Lambda_R/\Lambda_L)(\Lambda_{NL}/\Lambda_D)\gg 1$,
we again obtain the effective mass matrix (3.5) for the
active neutrinos $\nu_L$.

\noindent
{\bf  Case D}: \  $ M_L=0$

For a model without $\Phi_{NL}$, the matrix (3.1) leads to
$$
M^{6\times 6}\simeq  \left(
\begin{array}{cc}
m_L M_D^{-1} M_{R} M_D^{T-1} m_L^T & -m_L M_D^{-1} m_R \\
-m_R^T M_D^{T-1} m_L^T & 0
\end{array} \right) \ .
\eqno(3.12)
$$
When  $\Lambda_{NR}/\Lambda_{D} \ll \Lambda_R/\Lambda_L$,
the case gives a model with pseudo-Dirac neutrinos whose
masses are  given by (3.10).
When $\Lambda_{NR}/\Lambda_{D} \gg \Lambda_R/\Lambda_L$,
the case gives
$$
M(\nu_L) \simeq m_L M_D^{-1} M_R M_D^{T-1}  m_L^T  \ .
\eqno(3.13)
$$

Thus, in order to obtain  small values of the neutrino
masses ($m_\nu/m_{e,q} \sim 10^{-9}$), we need to seek for a model 
with $(\Lambda_R/\Lambda_L)
(\Lambda_R/\Lambda_L) \sim 10^9$, i.e., 
$$
x_1+x_3 \sim  9 \ .
\eqno(3.14)
$$

%%%%%%%%%%%%%%%%%%%%%%%%%%%%%%%%%%%%%%%%%%%%%%%%%%%%%%%%%%%%%%%%%
\vspace{.2in}

\section{Evolution of the gauge coupling constants}
\label{sec:4}

%\vglue.1in

For convenience, let us discuss on the case RLRL.
The electric charge operator $Q$ is given by
$$
Q=I_3^L +\frac{1}{2} Y \ ,
\ \ \ \ \Lambda_L < \mu \leq \Lambda_R\ ,
\eqno(4.1)
$$
$$
\frac{1}{2} Y=I_3^R + \frac{1}{2} Y_{LR} \ , 
\ \ \ \ \Lambda_R < \mu \leq \Lambda_F\ ,
\eqno(4.2)
$$
$$
\frac{1}{2} Y_{LR}= \frac{1}{2} Y_L +\frac{1}{2} Y_R \ ,
\ \ \ \ \Lambda_F < \mu \leq \Lambda_{NL}\ ,
\eqno(4.3)
$$
$$
\frac{1}{2} Y_L =I_{3}^{\prime L}+ \sqrt{\frac{2}{3}} F_{15}^L \ ,
\ \ \ \ \Lambda_{NL} < \mu \leq \Lambda_{XL}\ ,
\eqno(4.4)
$$
$$
\frac{1}{2} Y_R =I_{3}^{\prime R} +\sqrt{\frac{2}{3}} F_{15}^R \ ,
\ \ \ \ \Lambda_{NR} < \mu \leq \Lambda_{XR}\ ,
\eqno(4.5)
$$
where $I_3^{\prime L}$, $I_3^{\prime R}$, $F_{15}^L$ and $F_{15}^R$ 
are generators of ${\rm SU(2)}'_L$, ${\rm SU(2)}'_R$, SU(4)$_L$
and SU(4)$_R$, respectively.
We denote the gauge coupling constants corresponding to the
operators $Q$, $Y$, $Y_{LR}$, $Y_L$, $Y_R$, $I^L$, $I^R$, 
$I^{\prime L}$, $I^{\prime R}$, $F^L$ and $F^R$ as
$g_{em}\equiv e$, $g_1$, $g_{1LR}$, $g_{1L}$, $g_{1R}$, $g_{2L}$,
$g_{2R}$, $g'_{2L}$, $g'_{2R}$, $g_{4L}$ and $g_{4R}$, respectively.
The boundary conditions for these gauge coupling constants
at $\mu=\Lambda_L$, $\mu=\Lambda_R$, $\mu=\Lambda_F$,
$\mu=\Lambda_{NL}$, and $\mu=\Lambda_{NR}$ 
are as follows:
$$
\alpha^{-1}_{em}(\Lambda_L) = \alpha_{2L}^{-1}(\Lambda_L)+
\frac{5}{3}\alpha_1^{ -1}(\Lambda_L) \ ,
\eqno(4.6)
$$
$$
\frac{5}{3}\alpha^{ -1}_{1}(\Lambda_R) = 
\alpha_{2R}^{-1}(\Lambda_R)+
\frac{2}{3}\alpha_{1LR}^{-1}(\Lambda_R) \ ,
\eqno(4.7)
$$
$$
\frac{2}{3}\alpha^{-1}_{1LR}(\Lambda_F) = 
\frac{5}{3}\alpha^{-1}_{1L}(\Lambda_F)+
\frac{5}{3}\alpha_{1R}^{-1}(\Lambda_F) \ ,
\eqno(4.8)
$$
$$
\frac{5}{3}\alpha^{-1}_{1L}(\Lambda_{NL}) = 
\alpha^{\prime -1}_{2L}(\Lambda_{NL})
+\frac{2}{3}\alpha^{-1}_{4L}(\Lambda_{NL})\ ,
\eqno(4.9)
$$
and
$$
\frac{5}{3}\alpha^{-1}_{1R}(\Lambda_{NR}) = 
\alpha^{\prime -1}_{2R}(\Lambda_{NR})+
\frac{2}{3}\alpha_{4R}^{-1}(\Lambda_{NR}) \ ,
\eqno(4.10)
$$
respectively, correspondingly to Eqs.~(4.1) - (4.5),
where  $\alpha_i\equiv g_i^2/4\pi$. 
We also have the following boundary conditions at 
$\mu=\Lambda_F$, $\mu=\Lambda_{NL}$, $\mu=\Lambda_{NR}$,
$\mu=\Lambda_{XL}$  and $\mu=\Lambda_{XR}$:
$$
\alpha^{-1}_{3}(\Lambda_F) = \alpha^{-1}_{3L}(\Lambda_F)+
\alpha_{3R}^{-1}(\Lambda_F) \ ,
\eqno(4.11)
$$
$$
\alpha^{-1}_{3L}(\Lambda_{NL}) = \alpha^{-1}_{4L}(\Lambda_{NL})\ ,
\eqno(4.12)
$$
$$
\alpha^{-1}_{3R}(\Lambda_{NR}) = \alpha_{4R}^{-1}(\Lambda_{NR}) \ ,
\eqno(4.13)
$$
$$
\alpha^{-1}_{2L}(\Lambda_{XL}) = \alpha^{\prime -1}_{2L}
(\Lambda_{XL}) = \alpha_{4L}^{-1}(\Lambda_{XL}) \ ,
\eqno(4.14)
$$
$$
\alpha^{-1}_{2R}(\Lambda_{XR}) = \alpha^{\prime -1}_{2R}
(\Lambda_{XR}) = \alpha_{4R}^{-1}(\Lambda_{XR}) \ .
\eqno(4.15)
$$

The evolutions of the gauge coupling constants $g_i$ at
one-loop are given by the equations
$$
\frac{d}{dt} \alpha_i(\mu) = -\frac{1}{2\pi}
 b_i \alpha_i^2(\mu) \  , 
\eqno(4.16)
$$
where $t=\ln \mu$.

For example, for the case RLRL, the coefficients $b_i$ are 
calculated as follows.
The quantum numbers of the fermions $f$ and $F$ are 
assigned as those in Table \ref{T-qn}. 
Note that in the model with det$M_U=0$, the heavy fermions 
$F_L$ and $F_R$ except for $U_{3L}$ and $U_{3R}$ are decoupled 
for $\mu\leq\Lambda_F$ and the fermions $u_{3R}$ and $U_{3L}$ 
are decoupled for $\mu\leq\Lambda_R$.
Components of the Higgs scalars 
$\Phi_{NR}$, $\Phi_F$ and $\phi_R$ which contribute to
the coefficients $b_{iR}$, for example, in the energy-scale
range 6, are $[1; (1,3,\overline{10})+(3,1,10)]$, 
$[\overline{16}; (1,2,\overline{4})]$,
and $[1; (2,2,1)]$ of ${\rm SO(10)}_L\times 
[{\rm SU(2)}\times {\rm SU(2)}'\times {\rm SU(4)}]_R$,
respectively.
In the range 5, those become $[(1,1,1); (1,3,\overline{10})+
(3,1,10)]$, $[\overline{(1,2,\overline{4})}; (1,2,\overline{4})]$,
and $[(1,1,1); (2,2,1)]$ of 
$[{\rm SU(2)}\times {\rm SU(2)}'\times {\rm SU(4)}]_L \times
[{\rm SU(2)}\times {\rm SU(2)}'\times {\rm SU(4)}]_R$,
respectively.
The results are listed in Table \ref{T-b}.
The coefficients $b_i$ for the other cases can be
calculated in a similar way.

For the numerical study, we use the following input 
values \cite{pdg98}: 
$\alpha_1=0.01683$, $\alpha_2=0.03349$, and 
$\alpha_3=0.1189$  at $\mu=m_Z$ instead of  
those at $\mu=\Lambda_L$, and 
$$
\frac{\Lambda_R}{\Lambda_F}=0.02 \ , \ \ {\rm i.e.,}\ 
x_2=\log\frac{\Lambda_F}{\Lambda_R}=1.70\ ,
\eqno(4.17)
$$
which was derived from the observed ratio $m_c/m_t$ and
the modified universal seesaw model \cite{KFzp} with 
the constraint det$M_U=0$.
Since we have four constraint equations (4.14) and (4.15),
four of the eight parameters $x_1$, $x_3$, $x_4$, $x_5$, $x_6$ and
$$ 
a_1=\alpha_{1R}^{-1}(\Lambda_F)\ , \ \ 
a_2=\alpha_{2R}^{-1}(\Lambda_R)\ , \ \ 
a_3=\alpha_{3R}^{-1}(\Lambda_F)\ , 
\eqno(4.18)
$$
are independent. 
(For the cases with $x_4=0$, the number of the independent 
parameters are three.)
What is of great interest to us is whether we can give a 
reasonable order of the neutrino masses or not.
Therefore, we evaluate the maximal value of $x_3$ under the
constraints $x_1\geq 1$, $x_4\geq 0$, $x_5 \geq 0$, $x_6 \geq 0$,
$a_1\geq 1$, $a_2 \geq 1$ and $a_3 \geq 1$.
If there is no solution with $(x_3)_{max} \geq 0$, the case will be 
ruled out.  Even if we have a solution with $(x_3)_{max} \geq 0$,
which means that there are the unification points $\Lambda_{XL}$
and $\Lambda_{XR}$ of the gauge coupling constants,  
it will be difficult to explain the smallness of the neutrino 
masses if the numerical results of $x_1+x_3$ show $x_1+x_3<8$.

%%%%%%%%%%%%%%%%%%%%%%%%%%%%%%%%%%%%%%%%%%%%%%%%%%%%%%%%%%%%%%%%%
\vspace{.2in}

\section{Numerical results}
\label{sec:5}

%\vglue.1in

The numerical study has been done for non-SUSY and SUSY
cases with $n_F=1,2,3,4$ and $n_\phi=1,2,\cdots ,6$,
where $n_F$ and $n_\phi$ are numbers
of the Higgs scalars $\Phi_F$ and $\phi_L$ ($\phi_R$).

In all the non-SUSY cases with $\Lambda_{XR}>\Lambda_{XL}$, 
there is no solution with $(x_3)_{max}\geq 0$.
Therefore, the cases are ruled out.
For the non-SUSY case with  $\Lambda_{XL} \geq \Lambda_{XR}$, 
except for the case LRD which gives $(x_3)_{max}<0$, 
we can get positive values $(x_3)_{max}\geq 0$.
However, the solutions with $(x_3)_{max}\geq 0$ are allowed 
only when $n_F=1$ and $n_\phi \geq 5$. 
In Table \ref{T-nonSUSY}, the values of $(x_3)_{max}$ and
$(x_1)_{max}$ for the cases with $n_F=1$ and $n_\phi=6$ are
demonstrated.
As seen in Table \ref{T-nonSUSY}, these cases cannot give 
large values of $(x_1)_{max}$ and/or $(x_3)_{max}$.
Therefore, all the non-SUSY cases cannot explain the smallness 
of neutrino masses $m_\nu$, so that they are ruled out. 
(A similar study for a non-SUSY case, but for a case with different 
Higgs scalars, has been done by the author \cite{so10-epj}. 
In Ref.~\cite{so10-epj}, in spite of his numerical result 
$x_1 \leq 6.1$ for a case with $\Lambda_L >\Lambda_R$,
he concluded that  
the case  cannot be ruled out, because the numerical results 
should not be taken rigidly.
However, the discrepancy between $10^6$ and $10^9$ is too large
to reconcile.  )

In the SUSY cases, there are solutions with $(x_3)_{max}>0$
for the cases $n_\phi \geq 3$,
but they are allowed only when $x_4=x_5=x_6=0$ for the cases 
RLRL, RLLR, LRLR, LRRL, RRLL, LLRR, RLLD, LRLD, RLRD, LRRD, 
and only when $x_5=x_6=0$ for the cases RLD, LRD, RLDN, LRDN.
This means that 
$\Lambda_{XL}=\Lambda_{XR}=\Lambda_{NL}=\Lambda_{NR}
=\Lambda_D$,
and the symmetries 
${\rm SO(10)}_L\times {\rm SO(10)}_R$ are directly broken into
the symmetries 
$[{\rm SU(2)}\times{\rm U(1)}\times{\rm SU(3)}]_L \times 
[{\rm SU(2)}\times{\rm U(1)}\times{\rm SU(3)}]_R$ neither via
SU(4) nor SU(5).
As seen in Table \ref{T-b}, the results are independent of
the number of $\Phi_F$, $n_F$, for the cases with 
$x_4=x_5=x_6=0$.
All the cases RLRL, RLLR, $\cdots$, LRDN give the same
numerical results for the cases with the same value of
$n_\phi$.
The results are listed in Table \ref{T-SUSY}.
The value of $x_1+x_3$ for the case $n_\phi=3$ is 
somewhat large compared with a desirable value
$x_1+x_3 \sim 9$.
If we take the numerical results rigidly, the case $n_\phi=6$
is  favorable to explain $m_\nu/m_{e,q} \sim 10^{-9}$.

Numerical results for a typical SUSY case with $n_\phi=4$ is as follows:
$x_1=11.66$, $x_3=x_4=x_5=x_6=0$, i.e.,
$$
\Lambda_L=0.912 \times 10^2 \ {\rm GeV}\ , \ \ \ 
\Lambda_R=4.17 \times 10^{13} \ {\rm GeV}\ , \ \ \ 
\Lambda_F=\Lambda_X=2.08 \times 10^{15} \ {\rm GeV}\ , 
\eqno(5.1)
$$
$$
\alpha_{1R}^{-1}(\Lambda_F)=\alpha_{2R}^{-1}(\Lambda_F)=2.41 \ ,
\ \ \ \alpha_{3R}^{-1}(\Lambda_R)=3.66 \ .
\eqno(5.2)
$$
Since the numerical results are same for all the cases RLRL, RLLR, 
$\cdots$, the simplest choice will be to consider a model 
without Majorana mass terms.

%%%%%%%%%%%%%%%%%%%%%%%%%%%%%%%%%%%%%%%%%%%%%%%%%%%%

%%%%%%%%%%%%%%%%%%%%%%%%%%%%%%%%%%%%%%%%%%%%%%%%%%%%%%%%%%%%%%%%%
\vspace{.2in}

\section{Conclusion}
\label{sec:6}

%\vglue.1in

In conclusion, by using one-loop evolution equation for the
gauge coupling constants, we have investigated possible
intermediate mass scales $\Lambda_R$, $\Lambda_F$, 
$\Lambda_D$, $\Lambda_{NL}$, $\Lambda_{NR}$, and the 
unification scales  $\Lambda_{XL}$ and $\Lambda_{XR}$  
in the universal seesaw model with the gauge unification
SO(10)$_L\times$SO(10)$_R$. 
The evolution has systematically been investigated for all 
the cases in which 
the symmetries SO(10)$_L\times$SO(10)$_R$ are broken into
$[{\rm SU(2)}\times{\rm SU(2)'}\times{\rm SU(4)}]_L \times 
[{\rm SU(2)}\times{\rm SU(2)'}\times{\rm SU(4)}]_R$.
We have evaluated the maximum values of 
$x_3=\log(\Lambda_D/\Lambda_F)$ and 
$x_1=\log(\Lambda_R/\Lambda_L)$, because 
in most cases, the neutrino masses are suppressed compared
with the charged lepton and quark masses $m_{e,q}$ by
the factor $10^{-(x_1+x_3)}$ as stated in Sec.~\ref{sec:3}. 

We have found that all the cases cannot give a model with
$\Lambda_D\gg \Lambda_F$, i.e., we obtain, at most, 
$(x_3)_{max}=1.27$ for the SUSY model with $n_\phi=6$.
Therefore, models based on the cases B and D discussed in 
Sec.~\ref{sec:3} are ruled out.
The cases A and C and a model without Majorana mass terms
are our possible choices.

The non-SUSY model with $\Lambda_{XR}>\Lambda_{XL}$ is also
ruled out, because they cannot have the value $(x_3)_{max}\geq 0$.
Although the non-SUSY model with $\Lambda_{XL}>\Lambda_{XR}$ 
can have the value $(x_3)_{max}\geq 0$, the values of $x_1$ 
and $x_3$ are not sufficiently large to explain the smallness
of the neutrino masses.

In the SUSY cases, there are solutions with $(x_3)_{max}>0$,
but they are allowed only when $x_4=x_5=x_6=0$ and 
$n_\phi \geq 3$. 
The constraint $x_4=x_5=x_6=0$  means that the symmetries 
${\rm SO(10)}_L\times {\rm SO(10)}_R$ are directly broken into
the symmetries 
$[{\rm SU(2)}\times{\rm U(1)}\times{\rm SU(3)}]_L \times 
[{\rm SU(2)}\times{\rm U(1)}\times{\rm SU(3)}]_R$ neither via
SU(4) nor SU(5).

The constraint $n_\phi \geq 3$ is somewhat unwilling,
because the cases induce the flavor-changing neutral currents
(FCNC).
If we take the numerical results in Table \ref{T-SUSY} rigidly,
the case with $n_\phi=6$ (or $n_\phi=7$) is favorable.
However, the case with $n_\phi=6$ will fatally bring the
FCNC problem to us.
If we postpone the FCNC problem to the future, the case
with $n_\phi=6$ ($n_\phi^{up}=3$ and $n_\phi^{down}=3$) is
favorable, where $n_\phi^{up}$ and $n_\phi^{down}$ are 
the numbers of the SU(2) doublet Higgs scalars which couple
with up- and down-quark sectors, respectively.
Such a multi-Higgs model may play a role of the hierarchical mass 
structure within the family, i.e., $(m_e, m_\mu, m_\tau)$
(for example, see Ref.~\cite{yk-s3}).

On the other hand, if we take the FCNC problem seriously, 
we must take the case with $n_\phi=3$.
For example, if we consider a case with $n_\phi^{up}=2$ and 
$n_\phi^{down}=1$, the FCNC appear only in the up-quark sector, 
so that the damage form the FCNC will be reduced a little.
However, the case gives an over-suppression of the neutrino masses
because of $x_1+x_3\sim 12$.
If we want to take the case $n_\phi=3$, the excuse for the
over-suppression is as follows:
the present numerical results should not be taken too rigidly.
(i) The results were obtained by using the one-loop evolution 
equation (4.16).
The consideration of the higher order corrections may
slightly change the numerical results.
(ii) The constraint (3.14), i.e., $m_\nu \sim 10^{x_1+x_3}m_{e,q}$,
is obtained for the case that $y_L\sim 1$, $y_R\sim 1$ and
$y_F\sim 1$, where the Yukawa coupling constants $y_L$, $y_R$ and
$y_F$ are defined by $m_L=y_L \Lambda_L$, $m_R=y_R \Lambda_R$ and
$M_F=y_F \Lambda_F$, respectively.
For example, if we suppose the case that $y_L\sim 1$, $y_R\sim 1$ and
$y_F\sim 10^{-1}$, then the suppression can be reduced by a factor $10^1$.
(iii) We have used the numerical constraint 
$\Lambda_R/\Lambda_F=0.02$.
The constraint came from the phenomenological study \cite{KFzp}
 of the quark masses based on the universal seesaw model.
The value is model-dependent.
[However, it is likely that the value of $\Lambda_R/\Lambda_F$
is of the order of $m_c/m_t$ ($m_b/m_t$) in the framework of the 
new universal seesaw model.
Therefore, the numerical conclusion is still reliable as the order.]
Thus, we cannot completely exclude the case with $n_\phi=3$.

We conclude that when we intend to give a gauge unification
of the universal seesaw model, the SUSY 
${\rm SO(10)}_L\times {\rm SO(10)}_R$ mode is the most attractive
one, where the symmetries are directly broken into
$[{\rm SU(2)}\times{\rm U(1)}\times{\rm SU(3)}]_L \times 
[{\rm SU(2)}\times{\rm U(1)}\times{\rm SU(3)}]_R$ 
and the number of the weak doublet Higgs
scalars is $n_\phi \geq 3$, although the case still remains
problems.
It is our next task to investigate whether the Yukawa coupling
constants show reasonable behaviors under these symmetries or not.

%%%%%%%%%%%%%%%%%%%%%%%%%%%%%%%%%%%%%%%%%%%
%\vspace*{.2in}
\newpage

\centerline{\Large\bf Acknowledgments}

The author would like to thank H.~Fusaoka for  valuable suggestion
on the numerical estimates of $(x_3)_{max}$, T.~Fukuyama for 
bringing the importance of the SO(10) {\bf 126} Higgs scalar
to his attention, and M.~Tanimoto and N.~Okamura for their helpful
comments on the evolution equations.

%\newpage
%%%%%%%%%%%%%%%%
\vglue.2in

%%%%%%%%%%%%%%%%%
%%%%%%%%%%%%%%%%%
\newpage
%%%%%%%%%%%%%%%%%%%%%%%%%%%%%%%%%%%%%%

\begin{table}
\caption{Surviving symmetries in each energy-scale range 
and definition of the parameters $x_i$ for typical cases.
G$_{224}$ and G$_{123}$ denote 
${\rm G}_{224}={\rm SU(2)}\times {\rm SU(2)}' \times 
{\rm SU(4)}$ and ${\rm G}_{123}={\rm U(1)}\times {\rm SU(2)} 
\times {\rm SU(3)}$, respectively. 
For all cases, the symmetries in the ranges 1 and 2 are
given by ${\rm SU(2)}_L \times \times {\rm U(1)}_{Y}
\times {\rm SU(3)}_{LR}$ and 
${\rm SU(2)}_L \times {\rm SU(2)}_R \times {\rm U(1)}_{LR}
\times {\rm SU(3)}_{LR}$, respectively,
and the parameters $x_1$ and $x_2$ are defined by
$x_1=\log(\Lambda_R/\Lambda_L)$ and 
$x_2=\log(\Lambda_F/\Lambda_R)$.
For all cases, we can read $x_3$ as
$x_3=\log(\Lambda_{D}/\Lambda_F)$ 
\label{T-sym}}

\vglue.1in
\noindent
\begin{tabular}{|c|cccc|} \hline
Case & Range 3 & Range 4 & Range 5 & Range 6 \\ \hline
LRLR & $({\rm G}_{123})_L \times ({\rm G}_{123})_R$ &
$({\rm G}_{123})_L \times ({\rm G}_{224})_R$ &
$({\rm G}_{224})_L \times ({\rm G}_{224})_R$ &
$({\rm G}_{224})_L \times SO(10)_R$ \\
     & $x_3=\log(\Lambda_{NR}/\Lambda_F)$ &
$x_4=\log(\Lambda_{NL}/\Lambda_{NR})$ &
$x_5=\log(\Lambda_{XR}/\Lambda_{NL})$ &
$x_6=\log(\Lambda_{XL}/\Lambda_{XR})$ \\ \hline
LLRR & $({\rm G}_{123})_L \times ({\rm G}_{123})_R$ &
$({\rm G}_{123})_L \times ({\rm G}_{224})_R$ &
$({\rm G}_{123})_L \times SO(10)_R$ &
$({\rm G}_{224})_L \times SO(10)_R$ \\
     & $x_3=\log(\Lambda_{NR}/\Lambda_F)$ &
$x_4=\log(\Lambda_{XR}/\Lambda_{NR})$ &
$x_5=\log(\Lambda_{NL}/\Lambda_{XR})$ &
$x_6=\log(\Lambda_{XL}/\Lambda_{NL})$ \\ \hline
LRLD & $({\rm G}_{123})_L \times ({\rm G}_{123})_R$ &
$({\rm G}_{123})_L \times ({\rm G}_{224})_R$ &
$({\rm G}_{224})_L \times ({\rm G}_{224})_R$ &
$({\rm G}_{224})_L \times SO(10)_R$ \\
     & $x_3=\log(\Lambda_{D}/\Lambda_F)$ &
$x_4=\log(\Lambda_{NL}/\Lambda_{D})$ &
$x_5=\log(\Lambda_{XR}/\Lambda_{NL})$ &
$x_6=\log(\Lambda_{XL}/\Lambda_{XR})$ \\ \hline
LRRD & $({\rm G}_{123})_L \times ({\rm G}_{123})_R$ &
$({\rm G}_{224})_L \times ({\rm G}_{123})_R$ &
$({\rm G}_{224})_L \times ({\rm G}_{224})_R$ &
$({\rm G}_{224})_L \times SO(10)_R$ \\
     & $x_3=\log(\Lambda_{D}/\Lambda_F)$ &
$x_4=\log(\Lambda_{NR}/\Lambda_{D})$  &
$x_5=\log(\Lambda_{XR}/\Lambda_{NR})$ &
$x_6=\log(\Lambda_{XL}/\Lambda_{XR})$ \\ \hline
LRD & $({\rm G}_{123})_L \times ({\rm G}_{123})_R$ &
 &
$({\rm G}_{224})_L \times ({\rm G}_{224})_R$ &
$({\rm G}_{224})_L \times SO(10)_R$ \\
     & $x_3=\log(\Lambda_{D}/\Lambda_F)$ &
 &
$x_5=\log(\Lambda_{XR}/\Lambda_{D})$ &
$x_6=\log(\Lambda_{XL}/\Lambda_{XR})$ \\ \hline
LRDN & $({\rm G}_{123})_L \times ({\rm G}_{123})_R$ &
 &
$({\rm G}_{224})_L \times ({\rm G}_{224})_R$ &
$({\rm G}_{224})_L \times SO(10)_R$ \\
     & $x_3=\log(\Lambda_{D}/\Lambda_F)$ &
 &
$x_5=\log(\Lambda_{XR}/\Lambda_{D})$ &
$x_6=\log(\Lambda_{XL}/\Lambda_{XR})$ \\ \hline
\end{tabular}

\end{table}

%%%%%%%%%%%%%%%%%%%%%%%%%%%%%%%%%%%%%%%%%%%%%%%%%%%%%%%
%\newpage
%%%%%%%%%%%%%%%%%%%%%%%%%%%%%%%%%%%%%%

\begin{table}
\caption{Quantum numbers of the fermions $f$ and $F$ for
$[{\rm SU(2)}\times {\rm SU(2)}' \times {\rm SU(4)}]_L$
$\times$ $[{\rm SU(2)}\times {\rm SU(2)}' \times {\rm SU(4)}]_R$.
\label{T-qn}}

\vglue.1in
\begin{tabular}{|c|ccc|c|ccc|} \hline
  & $I_3^L$ & $I_3^{\prime L}$ & $\sqrt{\frac{2}{3}}F_{15}^L$ 
&  & $I_3^R$ & $I_3^{\prime R}$ & $\sqrt{\frac{2}{3}}F_{15}^R$ 
\\ \hline
$u_L$ & $+\frac{1}{2}$ & 0 & $+\frac{1}{6}$ & 
$u_R$ & $+\frac{1}{2}$ & 0 & $+\frac{1}{6}$ \\ 
$d_L$ & $-\frac{1}{2}$ & 0 & $+\frac{1}{6}$ & 
$d_R$ & $-\frac{1}{2}$ & 0 & $+\frac{1}{6}$ \\ \hline
$\nu_L$ & $+\frac{1}{2}$ & 0 & $-\frac{1}{2}$ & 
$\nu_R$ & $+\frac{1}{2}$ & 0 & $-\frac{1}{2}$  \\
$e_L$ & $-\frac{1}{2}$ & 0 & $-\frac{1}{2}$ & 
$e_R$ & $-\frac{1}{2}$ & 0 & $-\frac{1}{2}$ \\ \hline
$D_R^c$ & 0 & $+\frac{1}{2}$ & $-\frac{1}{6}$ & 
$D_L^c$ & 0 & $+\frac{1}{2}$ & $-\frac{1}{6}$  \\
$U_R^c$ & 0 & $-\frac{1}{2}$  & $-\frac{1}{6}$  & 
$U_L^c$ &  0 & $-\frac{1}{2}$  & $-\frac{1}{6}$  \\ \hline
$E_R^c$ & 0 & $+\frac{1}{2}$  & $+\frac{1}{2}$ & 
$E_L^c$ & 0 & $+\frac{1}{2}$  & $+\frac{1}{2}$  \\ 
$N_R^c$ & 0 & $-\frac{1}{2}$  & $+\frac{1}{2}$  & 
$N_L^c$ &  0 & $-\frac{1}{2}$  & $+\frac{1}{2}$  \\ \hline
\end{tabular}

\end{table}
%%%%%%%%%%%%%%%%%%%%%%%%

\begin{table}
\caption{Coefficients $b_{iL}$ and $b_{iR}$ of the 
gauge-coupling-constant renormalization group equations 
in the case RLRL. 
The coefficients $b_i$ without the indices $L$ or $R$,
except for $b_1$ and $b_3$ in the regions 1 and 2,  
denote $b_i \equiv b_{iL}=b_{iR}$.\label{T-b}}
\vglue.1in

\begin{tabular}{|c|c|c|c|}\hline
Range &  Non-SUSY  &  SUSY  &    \\ \hline
Range 1 &  $b_3=7$   & $b_3=-3$ &  \\ 
$\Lambda_L <\mu\leq\Lambda_R$ 
 & $b_2=\frac{10}{3} -\frac{1}{6}h_2$ & $b_2=-\frac{1}{2} h_2$ 
 & $h_2=n_\phi$ \\
 & $b_1=-\left( 4+\frac{1}{5} h_1\right)$ & 
 $b_1=-\left( 6+\frac{3}{5} h_1\right)$ & $h_1= \frac{1}{2} n_\phi$ 
\\ \hline
Range 2 & $b_3= \frac{19}{3}$  & $b_3= 2$ &   \\ 
$\Lambda_R <\mu\leq\Lambda_F$ 
 & $b_2=\frac{10}{3} -\frac{1}{6} h_2$ & $ b_2=-\frac{1}{2} h_2$
  & $h_2=n_\phi$ \\
 & $b_1=-\left( \frac{20}{3}+\frac{1}{2} h_1\right)$ & 
 $b_1=-\left( 10+\frac{3}{2} h_1\right)$ & $h_1=  n_\phi$ \\ \hline
Range 3 & $b_3= 7 - \frac{1}{6} h_3$  & $b_3= 3 -\frac{1}{2} h_3$ 
& $h_3=6$  \\ 
$\Lambda_{F} <\mu\leq\Lambda_{NL}$ 
& $b_2=\frac{10}{3} -\frac{1}{6}h_2$ & $b_2=-\frac{1}{2} h_2$ 
& $h_2=n_\phi$ \\
 & $b_1=-\left( 4+\frac{1}{5} h_1\right)$ & 
 $b_1=-\left( 6 +\frac{3}{5} h_1\right)$ & 
 $h_1= \frac{21}{4}+ \frac{1}{2} n_\phi$ \\ \hline
Range 4 & $b_{4L}= \frac{32}{3} - \frac{1}{6} h_{4L}$  & 
 $b_{4L}= 6 -\frac{1}{2} h_{4L}$ & $h_{4L}=36+16$  \\ 
$\Lambda_{NL} <\mu\leq\Lambda_{NR}$ 
 & $b_{2L}=\frac{10}{3}-\frac{1}{6} h_{2L}$ 
 & $b_{2L}=-\frac{1}{2} h_{2L}$ & $h_{2L} = 40+ 2 n_\phi$ \\
  & $b'_{2L}=\frac{10}{3}-\frac{1}{6} h'_{2L}$ 
 & $b'_{2L}=-\frac{1}{2} h'_{2L}$ & $h'_{2L} = 40+32+ 2 n_\phi$ \\
 & $b_{3R}= 7 - \frac{1}{6} h_{3R}$  & $b_{3R}= 3 -\frac{1}{2} h_{3R}$ 
 & $h_{3R}=16$  \\ 
 & $b_{2R}=\frac{10}{3} -\frac{1}{6}h_{2R}$ & $b_{2R}=-\frac{1}{2} h_{2R}$ 
 & $h_{2R}=n_\phi$ \\
 & $b_{1R}=-\left( 4+\frac{1}{5} h_{1R}\right)$ & 
 $b_{1R}=-\left( 6 +\frac{3}{5} h_{1R}\right)$ & 
 $h_{1R}= \frac{43}{3}+ \frac{1}{2} n_\phi$ \\ \hline
 Range 5 & $b_{4}= \frac{32}{3} - \frac{1}{6} h_{4}$  & 
 $b_{4}= 6 -\frac{1}{2} h_{4}$ & $h_{4}=36+16 n_F$  \\ 
$\Lambda_{NR} <\mu\leq\Lambda_{XL}$  
  & $b_{2}=\frac{10}{3}-\frac{1}{6} h_{2}$ 
 & $b_{2}=-\frac{1}{2} h_{2}$ & $h_{2} = 40+ 2 n_\phi$ \\
  & $b'_{2}=\frac{10}{3}-\frac{1}{6} h'_{2}$ 
 & $b'_{2}=-\frac{1}{2} h'_{2}$ & $h'_{2} = 40+32 n_F + 2 n_\phi$ 
\\ \hline
Range 6 &    &    &   \\
$\Lambda_{XL} <\mu\leq\Lambda_{XR}$ 
 & $b_{4R}= \frac{32}{3} - \frac{1}{6} h_{4R}$  & 
 $b_{4R}= 6 -\frac{1}{2} h_{4R}$ & $h_{4R}=36+32 n_F$  \\ 
  & $b_{2R}=\frac{10}{3}-\frac{1}{6} h_{2R}$ 
 & $b_{2R}=-\frac{1}{2} h_{2R}$ & $h_{2R} = 40+ 2 n_\phi$ \\
  & $b'_{2R}=\frac{10}{3}-\frac{1}{6} h'_{2R}$ 
 & $b'_{2R}=-\frac{1}{2} h'_{2R}$ & $h'_{2R} = 40+64 n_F + 2 n_\phi$ \\
 \hline
\end{tabular}

\end{table}

%%%%%%%%%%%%%%%%%%%%%%%%

\begin{table}
\caption{Maximal values of $x_3=\log(\Lambda_D/\Lambda_F)$ and
$x_1=\log(\Lambda_R/\Lambda_L)$ in the non-SUSY cases with 
$n_F=1$ and $n_\phi=6$.
\label{T-nonSUSY}}
\vglue.1in

\begin{tabular}{|c|cc|cc|}\hline
Case    & $(x_3)_{max}$ & at $x_1$ & $(x_1)_{max}$ & at $x_3$ \\
\hline
LRLR  & $+0.642$ & 1.00 & $+5.20$  & 0.00  \\
LRRL  & $+0.642$ & 1.00 & $+5.20$  & 0.00  \\
LLRR  & $+0.301$ & 1.00 & $+2.11$  & 0.00  \\
LRLD  & $+0.647$ & 1.00 & $+5.28$  & 0.00  \\
LRRD  & $+0.628$ & 1.00 & $+4.99$  & 0.00  \\
LRD   & $-0.049$ & 1.00 & $+0.44$  & 0.00  \\
LRDN  & $+0.642$ & 1.00 & $+5.20$  & 0.00 \\ \hline
\end{tabular}

\end{table}

%%%%%%%%%%%%%%%%%%%%%%%%
%%%%%%%%%%%%%%%%%%%%%%%%

\begin{table}
\caption{Maximal values of $x_3=\log(\Lambda_D/\Lambda_F)$ and
$x_1=\log(\Lambda_R/\Lambda_L)$ in the SUSY cases with 
$n_\phi=3,4,6$.
\label{T-SUSY}}
\vglue.1in

\begin{tabular}{|c|cc|cc|c|}\hline
$n_\phi$  & $(x_3)_{max}$ & at $x_1$ & $(x_1)_{max}$ & at $x_3$ &
$x_1+x_3$ \\
\hline
3  & $+0.063$ & 12.27 & $+12.38$  & 0.00 & $12.38\geq x_1+x_3\geq 12.33$ \\
4  & $+0.570$ & 10.66 & $+11.66$  & 0.00 & $11.66\geq x_1+x_3\geq 11.23$ \\
6  & $+1.269$ & 8.16 & $+10.41$  & 0.00 & $10.41\geq x_1+x_3\geq 9.43$ \\
 \hline
\end{tabular}

\end{table}

%%%%%%%%%%%%%%%%%%%%%%%%

\end{document}